\documentclass[runningheads]{llncs}
\usepackage{graphicx}
\usepackage{amsmath}
\usepackage{amssymb}
\usepackage{optidef}
\usepackage{todonotes}
\usepackage{hyperref}
\usepackage{multirow}
\usepackage{diagbox}
\usepackage{adjustbox}
\usepackage{makecell}
\usepackage{comment}

\begin{document}
\title{FakeClaim: A Multiple Platform-driven Dataset for Identification of Fake News on 2023 Israel-Hamas War}

%
\titlerunning{FakeClaim: Fake News on Israel-Hamas War}
%
\author{Gautam Kishore Shahi\inst{1}\orcidID{0000-0001-6168-0132} \and
Amit Kumar Jaiswal\inst{2}\orcidID{0000-0001-8848-7041} \and
Thomas Mandl\inst{3}\orcidID{0000-0002-8398-9699}
}
\authorrunning{Shahi et al.}
%
\institute{University of Duisburg-Essen, Duisburg, Germany
\and
University of Surrey, Guildford, United Kingdom
\and
University of Hildesheim, Germany
\\
\email{gautam.shahi@uni-due.de, a.jaiswal@surrey.ac.uk, mandl@uni-hildesheim.de}
}
\maketitle              
%

\begin{abstract}

We contribute the first publicly available dataset of factual claims from different platforms and fake YouTube videos on the 2023 Israel-Hamas war for automatic fake YouTube video classification. The FakeClaim data is collected from 60 fact-checking organizations in 30 languages and enriched with metadata from the fact-checking organizations curated by trained journalists specialized in fact-checking. Further, we classify fake videos within the subset of YouTube videos using textual information and user comments. We used a pre-trained model to classify each video with different feature combinations. Our best-performing fine-tuned language model, Universal Sentence Encoder (USE), achieves a Macro F1 of 87\%, which shows that the trained model can be helpful for debunking fake videos using the comments from the user discussion. The dataset is available on Github\footnote{https://github.com/Gautamshahi/FakeClaim}.

\end{abstract}
%
%
%

\section{Introduction}


Social media are a source of much unreliable information. During conflicts, the percentage of misinformation often increases \cite{lovelace2022tomorrow,shahi2020fakecovid} and, for example, false statements or misleading visual content are shared over different platforms. Fact-checking by experts has proven to help debunk misinformation and support a rational discourse. Debunking of claims propagated on social media platforms such as \textsc{X}\footnote{We refer \textsc{X} as Twitter for clarity.} (formerly Twitter), YouTube, and Facebook \cite{shahi2021amused} has increased in the last decade. 

The fake news discourse changes frequently and follows current topics, like COVID-19, the Russia-Ukraine War, the 2023 Turkey-Syria earthquakes, and the 2023 Israel-Hamas war. Each time fake news spreads, it also makes use of new technology and patterns. The emergence of generative Artificial Intelligence (AI) led to fake news consisting of voice clips and altered media, which is hard to debunk. In contrast, ChatGPT is not able to detect fake claims without training data. For scientific analysis of the information space regarding conflicts from diverse angles, as well as for improving technology for supporting fact-checking, datasets are essential. We applied the AMUSED framework \cite{shahi2021amused} to collect a dataset for fake claims on the Israel-Hamas conflict, which broke out in early October 2023. This dataset will allow research to analyze the information space regarding the war from diverse angles.

\textbf{Contributions} We introduce the first fact-checked collection of claims related to the 2023 Israel-Hamas war called FakeClaim. It comprises 1,499 claims
collected from 1,370 fact-checked articles published by 60 fact-checking organizations in 30 languages with background information about evidence pages to verify or reject the claims. We perform content extraction for the fact-checked articles, such as the origin of the claim on social media or unreliable news sources, and extract them as sample data as shown in table~\ref{data:example}. We demonstrate the utility of the dataset by forming a YouTube video classification by training state-of-the-art prediction models and find that user engagement, comments on the video, evidence pages and metadata significantly contribute to model performance. Finally, we benchmark our data with a fine-tuned version of the pre-trained embedding model, namely, Universal sentence encoder~\cite{cer2018universal} that jointly ranks evidence pages and performs veracity prediction.
The best-performing model achieves a Macro F1 of 87\%, showing that combining these features can be used for video classification in future work.

\begin{table}[htb]
	\centering
	\caption{An example of a claim debunked by a fact-checking website and YouTube Video Link extraction from fact-checked articles (Claim data if extraction is possible like YouTube). }
\begin{tabular}{|p{2.7 cm} |p{9cm}|} 
\hline
\textbf{Feature} & \textbf{Value} \\ \hline
ClaimId & factly\_6  \\
Claim & Visuals of Israelis protesting against Prime Minister Benjamin Netanyahu for war against Palestine.   \\
Label & False  \\
Claim URL & \url{https://tinyurl.com/25vm9h2e}  \\
Article Title & March 2023 Video Shared as Recent Israelis Protest Against Netanyahu Amid Ongoing Israel-Hamas Conflict  \\
Article text & Amid the ongoing Israel-Hamas conflict that began in October 2023, a video depicting a massive crowd .... (refer to above link)  \\
Published date & October 22, 2023  \\
Language & English  \\
Claim Source & YouTube  \\
Link to Claim & \url{https://www.youtube.com/watch?v=XDJP5Ow3ri} \\
Claim(video) Data & Details extracted from claim(video)* \\
\hline
	\end{tabular}%
	\label{data:example}%
\end{table}

\section{Related Work}

Several datasets are available for recent conflicts and fake news from multiple social media platforms~\cite{shahi2021amused}. The analysis of public discussions on controversial issues has led to much research. 
A dataset for the Syrian war was assembled to support machine learning for fake news detection. The authors used the documentation of an NGO as ground truth~\cite{DBLP:conf/icwsm/SalemFEJF19}.

For the Russian invasion of Ukraine, a dataset from Twitter was collected using related hashtags. An analysis showed that the activities of government sponsored actors were high at the beginning of the conflict~\cite{DBLP:conf/icwsm/ChenF23}. A similar dataset for Facebook and Twitter shows how few actors contribute spread misinformation very intensely~\cite{DBLP:conf/websci/0002LJF23}. Another dataset was collected from Reddit and distinguished between military and conflict-related posts~\cite{zhu2022reddit}. A study using topic modeling for a dataset on the Ukraine war has shown that identifying false news is unfeasible using unsupervised methods~\cite{shin7content}.

Technology for detecting false news articles is often based on supervised machine learning. In the shared task, CheckThat! which was organized in 2022, the classification of the veracity of text content proved to be hard for systems, and the best F1 measure obtained was 0.33 for a four-class problem~\cite{DBLP:conf/clef/KohlerSSWS0S22,shahi2021overview}. Previously, a YouTube video classification by textual information was performed using BERT~\cite{devlin2019bert} without considering the evidence and user engagement~\cite{rochert2021networked}. Fake news classification has also been carried out for news articles using social user engagement~\cite{wu2023decor}. 

There is a lack of datasets as combinations of claims, fake posts (like videos) and background evidence. Till now, Fake video detection has not yet been performed using a combination of claims, textual information, and user engagement. We define the task  in section~\ref{sec:pf} and present the implementation of our system and the results in section~\ref{exp}.    


\section{Methodology}

In this section, we explain the data collection process, formally define the fake news detection task for posts (videos) and explain the problem formation.


\subsection{Data Collection}



For data collection, we selected a list of fact-checking organizations publishing content on the 2023 Israel-Hamas conflict, such as Snopes, BoomLive. The AMUSED framework~\cite{shahi2021amused} for annotation was applied. It allows the extraction of the claimed source from either a social media post or from an online news source and maps the content to the label or verdict of the fact-checked article used in studying Twitter~\cite{shahi2021exploratory}, YouTube~\cite{rochert2021networked}. 

First, we collect fact-checked articles from multiple fact-checking organizations. Starting from each collected article, we extract data available and translate it in case it is not written in English. Next, the content of the articles was filtered with words related to the Israel-Hamas war using keywords like \textit{Israel, Hamas}. We fetched social media links from filtered, fact-checked articles like YouTube videos and Instagram posts. Overall, we collected 1,370 fact-checking articles from 62 Fact-checking organizations in 30 languages. These fact-checked articles debunk 1,499 claims from different social media platforms. Due to the availability of the research API and the possibility of extracting comments as an expression of user engagement, we perform a classification of the subset of YouTube videos. We obtained 404 fake YouTube videos, of which 388 still remain online on YouTube (checked on 29-10-2023). We randomly filtered 610 videos from YouTube using a keyword search related to the Israel-Hamas war to obtain videos with most likely correct information as another class.


\begin{figure}
    \centering
    \includegraphics[width=.99\textwidth]{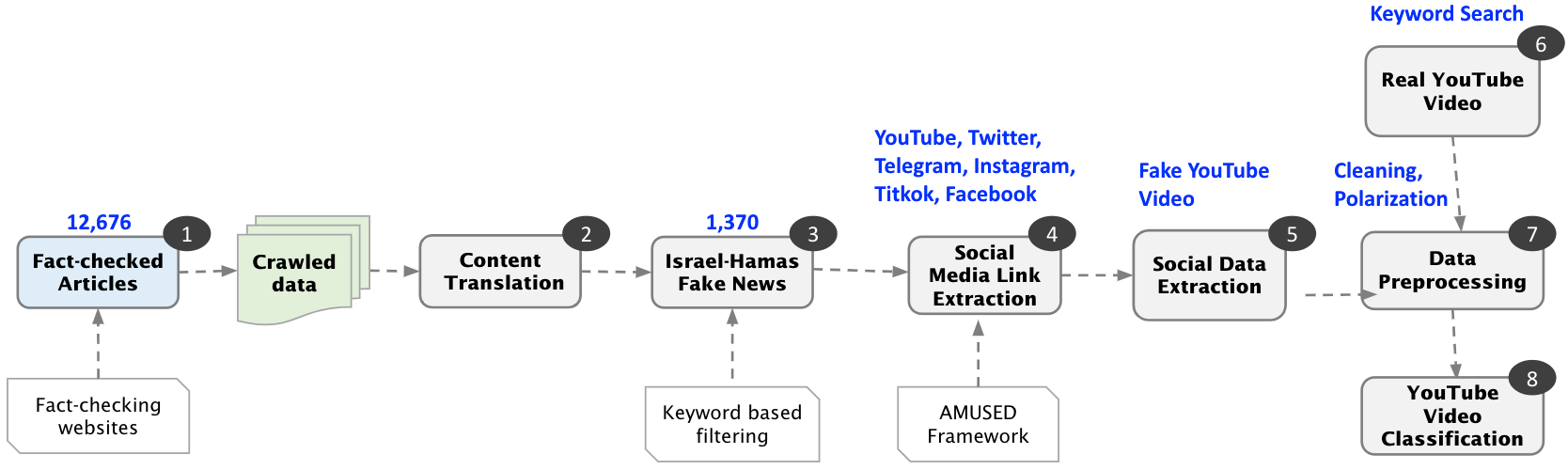}
    \caption{The research framework used for extracting the dataset and classification problem.}
    \label{figure:method}
\end{figure}


\begin{table}[tb]
	\centering
	\caption{Distribution of claim debunked from different social media platforms(* some fact-checked article debunks multiple claims from different platforms) }
\begin{tabular}{|p{2.5 cm} |p{3cm}|} 
\hline
\textbf{Claim Source} & \textbf{Number of Claims} \\ \hline
Facebook & 579 \\
Twitter & 183 \\
YouTube & 389 \\
Tiktok & 186 \\
Instagram & 90 \\
Telegram & 72 \\
\hline
Total(*)  & 1,499 \\
\hline
	\end{tabular}%
	\label{data:claims}%
\end{table}

\subsection{Problem Formulation}
\label{sec:pf}

Let D be a fake news detection dataset containing N samples. In the social media setting, we define the dataset as
                        \begin{equation*}
                 \mathcal{D} = \{\mathcal{V}, \mathcal{U}, \mathcal{C}, \mathcal{F}\} 
                        \end{equation*}
where $\mathcal{V} = \{v_1,v_2,v_3,....,v_n \}$ is a set of textual descriptions of \textbf{YouTube videos}, $\mathcal{U} =\{u_1,u_2,u_3,....,u_n \}$ is a set of related \textbf{social users} engaged in conversation related to the corresponding YouTube videos. $\mathcal{C}$ represent \textbf{user's comments} on videos, in which $c\in \mathcal{C}$ and $\mathcal{F}$ depicts the claim and provides correct background information to debunk the claim which is extracted from \textbf{fact-checked article},$f \in \mathcal{F}$, R represents the set of \textbf{social audience engagements and background truth} as the composition of ${\mathcal{V},\mathcal{U},\mathcal{C}}$ in which $r \in R $ is defined as a quadruplet  $\{(v,u,c,f)| v\in V, u \in U, c \in C, f \in F\}$ (i.e. user u has given c comments on YouTube video V and f was background truth provided by fact-checkers).

In line with most existing studies, we treat fake news detection on social media given the current scenario of YouTube framed as a binary classification problem (such as \cite{DBLP:conf/smsociety/AjaoBZ18,rochert2021networked}). Specifically, V is split into a training set V\textsubscript{train}
and test set V\textsubscript{test}. Each video v, which belongs to V\textsubscript{train}, is associated with a ground truth label y of 1 if V is fake and 0 otherwise. We formulate the problem as follows:

Problem definition (Fake News Detection on YouTube Video): Given a YouTube video dataset D = {V, U, C, F} and ground-truth training labels
y\textsubscript{train}, the goal is to obtain a classifier f that, given test videos V\textsubscript{train}, is able to predict the corresponding veracity labels y\textsubscript{test}.


\section{Experiments \& Results}
\label{exp}

\textbf{Data Prepossessing:} After collection, articles are preprocessed by removing special characters, URLs, and emoji and applying a state-of-the-art classification model. The research framework is explained in Figure~\ref{figure:method}. An overview of claims debunked from multiple platforms is shown in table~\ref{data:example}. The collected data is  openly available for further studies. Overall, after data cleaning and preprocessing steps which excluded videos without any comments, the dataset consists of 756 unique videos and 166,645 comments, with 301 videos of the fake category with 85,228 comments and 455 of the true category with 81,417 comments. 
After preprocessing, we computed frequent words using TF-IDF and then used textblob\footnote{https://textblob.readthedocs.io/en/dev/quickstart.html} to obtain the sentiment polarity for each comment as values in [-1,+1]. These can also be aggregated for each video.

\subsection{Experiment} 
Given our task of identifying fake claims, we utilize pre-trained word embedding models, where the multitude of textual comments spanned among YouTube videos entail diverse opinions, and so the social behaviors tend to upsurge conflict. We employ the polarity values for each video's comments as an implicit feature that distinguishes the spectrum of opinions reflecting online social interaction. We split our data by stratifying the binary class (fake/real claims) in the proportion of 70\% for the train set and 30\% for the test set and used precision (P), recall (R), and F1-score (F1) for evaluation. \\ 

\textit{Baseline Models:} Given our collected corpus of fact claims data, including comments and metadata, our task tends to word-/sentence-level classification, and so we utilize pre-trained word embeddings and fine-tune the base model, namely, GNews-Swivel-20D~\cite{shazeer2016swivel}. The following baselines were implemented: \\

\textit{GN-Swivel-20D:} The approach uses Swivel embeddings~\cite{shazeer2016swivel}, which result from training on the extensive Google News 130 GB corpus. The GN-Swivel-20 dimensional model transforms words or phrases into 20-dimensional vectors, thereby adeptly encapsulating the semantic connotations and contextual interconnections that underlie the vocabulary, as discerned from an extensive textual corpus. 

\textit{Universal Sentence Encoder~(USE):} We utilize the pre-trained Universal Sentence Encoder~\cite{cer2018universal} tailored to transform sentences and text paragraphs into high-dimensional vector representations. It captures encapsulates the semantic essence and contextual information of complete sentences.

\textit{RoBERTa~:} RoBERTa is an encoder-only transformer model with 125 million parameters which underwent training on an extensive dataset comprising 160 GB of text corpus~\cite{liu2019roberta}. Our selection of this model is motivated by its common usage in classification and regression tasks, a practice shared with other encoder-only models such as BERT~\cite{devlin2019bert,bommasani2021opportunities,jaiswal2023lightweight}. We employ the pre-trained RoBERTa$_{\text{base}}$ model\footnote{https://huggingface.co/roberta-base} for comparison.

\textbf{Fine-tuning Settings:} Initially, we set the learning rate of both GN-Swivel-20D and USE models to be 1$e$-3 with Adam as an optimizer and early stopping for monitoring the minimum validation loss and fine-tune the first layer for over ten epochs. Also, the model parameters are slightly smaller ($\sim$421k) in comparison to other variants of pre-trained embedding models. For fine-tuning the Universal Sentence Encoder model, we set the embedding size to 512, and then fine-tune the first layer. The model has $\sim$257 million parameters. We employ the pre-trained RoBERTa  model without fine-tuning due to limited availability of resources.

\textbf{Analysis:}
We outline how our results reported in Table~\ref{tab:tab_1} can reflect on mitigating factors for reducing fake claims in social discourse platforms. 
In Table~\ref{tab:tab_1}, `Video' refers to the video title. For this approach, we fed our model with concatenated pairs of video titles spanning comments and fact claims. We found out that the addition of fact-checked articles boosts the precision and F1 scores. On the contrary, we score the polarization for each fact-checked article, where most articles and comments are highly skewed toward the words `Israel,' `Hamas,' and `Gaza'. However, due to the skewness of the distribution of the sentiments for the comments tending to be high, our model performance fluctuates in the range of 72\%~-~74\%. Real videos are used without any fact-checked articles. Our current results are obtained without any additional polarisation features. This shows that the proliferation of users' engagement during discourse is an important factor in identifying fake claims.

\begin{table*}
    \centering
    \caption{Results with pre-trained word embedding models for classification of Fake claims. Model performance is based on the F1 score, best models are in bold.}
    \begin{adjustbox}{width=\linewidth}
    \begin{tabular}{|c|c|c|c|c|c|c|c|c|c|c|c|c|}\cline{3-13}
    \hline
    \multirow{4}{*}{\textbf{\diagbox{Model}{\thead{Class \& \\~~~Features}}}} &
    \multicolumn{6}{|c}{\textbf{fake (0)}} & \multicolumn{6}{|c|}{\textbf{real (1)}} \\ \cline{2-13}
     & \multicolumn{3}{|c}{\textbf{\thead{Video+\\Comments+\\ Claims}}} & %
        \multicolumn{3}{|c}{\textbf{\thead{Video+\\Comments}}} & \multicolumn{3}{|c|}{\textbf{\thead{Video+\\Comments+\\ Claims}}} & \multicolumn{3}{|c|}{\textbf{\thead{Video+\\Comments}}} \\ \cline{2-13}
    & \multicolumn{12}{c|}{Scores}\\
    \cline{2-13}
    & P & R & F1 & P & R & F1 & P & R & F1 & P & R & F1 \\
    \hline
    GN-Swivel-20D & 0.831 & 0.814 & 0.822 & 0.748 & 0.772 & 0.76 & 0.822 & 0.828 & 0.825 & 0.768 & 0.77 & 0.769 \\
    \hline
    Universal Sentence Encoder & 0.874 & 0.867 & \textbf{0.87} & 0.803 &  0.798 & \textbf{0.80} & 0.868 & 0.882 & \textbf{0.875} & 0.796 & 0.813 & \textbf{0.804} \\
    \hline
    RoBERTa & 0.69 & 0.75 & 0.72 & 0.66 & 0.72 & 0.69 & 0.64 & 0.69 & 0.664 & 0.70 & 0.69 & 0.689 \\
    \hline
    \end{tabular}
    \end{adjustbox}
    \label{tab:tab_1}
\end{table*}

\vspace{-4mm}
\section{Conclusion \& Future Work}
\vspace{-2mm}
In this study, we present a timely analysis of the 2023 Israel-Hamas war, providing a FakeClaim dataset that refers to the original fake posts and further scrapped YouTube videos and comments for the analysis. We formulated a classification problem for fake video descriptions by combining text information from videos, comments, claims and evidence and found that combining different features improved the F1 score. The study can be useful in fighting fake videos and help to mitigate fake news.

The limitations include the issue that the true class was not manually checked or fact-checked; in future work, we will try to also obtain labels for correct videos. Another limitation of the study is that collecting and extracting claims from fact-checked articles is difficult due to the lack of common standards for publishing fact-checked articles. It required continuous monitoring of the different websites and their structure to gather data. Finding around 1,499 claims in 3 weeks of conflict shows that the volume of claims is high, and an automated approach can help to debunk fake posts. The claims are spread over multiple platforms like YouTube, Twitter and Facebook. In future work, we intend to collect, explore and investigate FakeClaim data given the diversity of online comments from social platforms such as TikTok.

\section{FAIR Data and Ethical Considerations}
We have conducted all experiments on a macro level following
strict data access, storage, and auditing procedures for the sake of
accountability. Users personal information (like author name, profile picture) is neither used nor stored for classification in this paper 
The FakeClaim dataset, together with the ML/DL models generated in this study is available publicly for further research following the policy for sharing data.

\bibliographystyle{unsrt}
\bibliography{israelwar}

\begin{thebibliography}{10}

\bibitem{lovelace2022tomorrow}
Alexander~G Lovelace.
\newblock Tomorrow’s wars and the media.
\newblock {\em The US Army War College Quarterly: Parameters}, 52(2):117--134, 2022.

\bibitem{shahi2020fakecovid}
Gautam~Kishore Shahi and Durgesh Nandini.
\newblock {FakeCovid--A multilingual cross-domain fact check news dataset for COVID-19}.
\newblock {\em arXiv preprint arXiv:2006.11343}, 2020.

\bibitem{shahi2021amused}
Gautam~Kishore Shahi and Tim~A Majchrzak.
\newblock Amused: an annotation framework of multimodal social media data.
\newblock In {\em International Conference on Intelligent Technologies and Applications}, pages 287--299. Springer, 2021.

\bibitem{cer2018universal}
Daniel Cer, Yinfei Yang, Sheng{-}yi Kong, Nan Hua, Nicole Limtiaco, Rhomni~St. John, Noah Constant, Mario Guajardo{-}Cespedes, Steve Yuan, Chris Tar, Brian Strope, and Ray Kurzweil.
\newblock {Universal Sentence Encoder for English}.
\newblock In {\em Proceedings of the Conference on Empirical Methods in Natural Language Processing, {EMNLP}: System Demonstrations, Brussels, Belgium, October 31 - November 4,}, pages 169--174. Association for Computational Linguistics, 2018.

\bibitem{DBLP:conf/icwsm/SalemFEJF19}
Fatima K.~Abu Salem, Roaa~Al Feel, Shady Elbassuoni, Mohamad Jaber, and May Farah.
\newblock {FA-KES:} {A} fake news dataset around the {Syrian} war.
\newblock In {\em Proceedings of the Thirteenth International Conference on Web and Social Media, {ICWSM} 2019, Munich, Germany, June 11-14, 2019}, pages 573--582. {AAAI} Press, 2019.

\bibitem{DBLP:conf/icwsm/ChenF23}
Emily Chen and Emilio Ferrara.
\newblock Tweets in time of conflict: {A} public dataset tracking the {Twitter Discourse on the War between Ukraine and Russia}.
\newblock In {\em Proceedings of the Seventeenth International {AAAI} Conference on Web and Social Media, {ICWSM} 2023, June 5-8, 2023, Limassol, Cyprus}, pages 1006--1013. {AAAI} Press, 2023.

\bibitem{DBLP:conf/websci/0002LJF23}
Francesco Pierri, Luca Luceri, Nikhil Jindal, and Emilio Ferrara.
\newblock Propaganda and misinformation on {Facebook and Twitter during the Russian Invasion of Ukraine}.
\newblock In {\em Proceedings of the 15th {ACM} Web Science Conference, WebSci , Austin, TX, USA, 30 April 2023 - 1 May}, pages 65--74. {ACM}, 2023.

\bibitem{zhu2022reddit}
Yiming Zhu, Ehsan-ul Haq, Lik-Hang Lee, Gareth Tyson, and Pan Hui.
\newblock {A Reddit dataset for the Russo-Ukrainian conflict in 2022}.
\newblock {\em arXiv preprint arXiv:2206.05107}, 2022.

\bibitem{shin7content}
Yucheol Shin, Yvan Sojdehei, Limin Zheng, and Brad Blanchard.
\newblock {Content-Based Unsupervised Fake News Detection on Ukraine-Russia War}.
\newblock {\em SMU Data Science Review}, 7(1):3.

\bibitem{DBLP:conf/clef/KohlerSSWS0S22}
Juliane K{\"{o}}hler, Gautam~Kishore Shahi, Julia~Maria Stru{\ss}, Michael Wiegand, Melanie Siegel, Thomas Mandl, and Mina Sch{\"{u}}tz.
\newblock Overview of the {CLEF-2022} checkthat! lab: Task 3 on fake news detection.
\newblock In {\em Proceedings of the Working Notes of {CLEF} 2022 - Conference and Labs of the Evaluation Forum, Bologna, Italy, Sept. 5th-8th}, pages 404--421. CEUR-WS.org, 2022.

\bibitem{shahi2021overview}
Gautam~Kishore Shahi, Julia~Maria Stru{\ss}, and Thomas Mandl.
\newblock Overview of the {CLEF-2021} checkthat! lab: Task 3 on fake news detection.
\newblock In {\em Proceedings of the Working Notes of {CLEF} - Conference and Labs of the Evaluation Forum, Bucharest, Romania, September 21st - to - 24th.}, volume 2936 of {\em {CEUR} Workshop Proceedings}, pages 406--423. CEUR-WS.org, 2021.

\bibitem{devlin2019bert}
Jacob Devlin, Ming-Wei Chang, Kenton Lee, and Kristina Toutanova.
\newblock Bert: Pre-training of deep bidirectional transformers for language understanding.
\newblock In {\em Proceedings of the 2019 Conference of the North American Chapter of the Association for Computational Linguistics: Human Language Technologies, Volume 1 (Long and Short Papers)}, pages 4171--4186, 2019.

\bibitem{rochert2021networked}
Daniel R{\"{o}}chert, Gautam~Kishore Shahi, German Neubaum, Bj{\"{o}}rn Ross, and Stefan Stieglitz.
\newblock The networked context of {COVID-19} misinformation: Informational homogeneity on youtube at the beginning of the pandemic.
\newblock {\em Online Soc. Networks Media}, 26:100164, 2021.

\bibitem{wu2023decor}
Jiaying Wu and Bryan Hooi.
\newblock Decor: Degree-corrected social graph refinement for fake news detection.
\newblock In {\em Proceedings of the 29th ACM SIGKDD Conference on Knowledge Discovery and Data Mining}, pages 2582--2593, 2023.

\bibitem{shahi2021exploratory}
Gautam~Kishore Shahi, Anne Dirkson, and Tim~A Majchrzak.
\newblock {An exploratory study of COVID-19 misinformation on Twitter}.
\newblock {\em Online Social Networks and Media}, 22:100104, 2021.

\bibitem{DBLP:conf/smsociety/AjaoBZ18}
Oluwaseun Ajao, Deepayan Bhowmik, and Shahrzad Zargari.
\newblock Fake news identification on {Twitter} with hybrid {CNN} and {RNN} models.
\newblock In {\em Proceedings of the 9th International Conference on Social Media and Society, SMSociety 2018, Copenhagen, Denmark, July 18-20, 2018}, pages 226--230. {ACM}, 2018.

\bibitem{shazeer2016swivel}
Noam Shazeer, Ryan Doherty, Colin Evans, and Chris Waterson.
\newblock Swivel: Improving embeddings by noticing what's missing.
\newblock {\em arXiv preprint arXiv:1602.02215}, 2016.

\bibitem{liu2019roberta}
Yinhan Liu, Myle Ott, Naman Goyal, Jingfei Du, Mandar Joshi, Danqi Chen, Omer Levy, Mike Lewis, Luke Zettlemoyer, and Veselin Stoyanov.
\newblock Roberta: A robustly optimized bert pretraining approach.
\newblock {\em arXiv preprint arXiv:1907.11692}, 2019.

\bibitem{bommasani2021opportunities}
Rishi Bommasani, Drew~A Hudson, Ehsan Adeli, Russ Altman, Simran Arora, Sydney von Arx, Michael~S Bernstein, Jeannette Bohg, Antoine Bosselut, Emma Brunskill, et~al.
\newblock On the opportunities and risks of foundation models.
\newblock {\em arXiv preprint arXiv:2108.07258}, 2021.

\bibitem{jaiswal2023lightweight}
Amit~Kumar Jaiswal and Haiming Liu.
\newblock Lightweight adaptation of neural language models via subspace embedding.
\newblock In {\em Proceedings of the 32nd ACM International Conference on Information and Knowledge Management}, pages 3968--3972, 2023.

\end{thebibliography}

\end{document}